\documentclass[preprint,showpacs,preprintnumbers,amsmath,amssymb]{revtex4}

\usepackage{graphicx}
\usepackage{dcolumn}
\usepackage{bm}

\newcommand{\comm}[2]{[#1,#2]}

\newcommand{\trace}{{\rm Tr}}

\newcommand{\braketo}[3]{\langle #1 | #2 | #3 \rangle}

\newcommand{\ket}[1]{| #1 \rangle}
\newcommand{\lbar}[1]{\overline{#1}}

\begin{document}

\preprint{}

\title{Stability of the rotating SU(3) Skyrmion}

\author{Satoru Akiyama}
\email{akiyama@ph.noda.tus.ac.jp}

\author{Masahiro Kawabata}

\affiliation{
	Department of Physics,
	Faculty of Science and Technology,
	Tokyo University of Science,
	2641, Noda, Chiba 278-8510, Japan 
}

\date{\today}

\begin{abstract}
The profile functions of the SU(3) Skyrme soliton are investigated
for the octet, decuplet, and antidecuplet baryons by the mean field approach.
In this approach, the profile functions are affected by the spatial rotation, the flavor rotation,
and the flavor symmetry breaking.
The solitons are stable only in the restricted areas of the parameter space for each multiplet.
When the flavor symmetry breaking is large,
the area for the antidecuplet is narrow compared to those for the octet and decuplet.
The parameters are determined by the baryon mass spectrum, and
the deformation of the soliton has sizable effects on the masses.
\end{abstract}

\maketitle

\section{Introduction\label{sec:intro}}
Diakonov, Petrov, and Polyakov~\cite{rf:Diakonov97} made
a detailed prediction for the masses and the decay widths of the antidecuplet ($\lbar{\bf 10}$) baryons
in the framework of the Skyrme soliton (Skyrmion) model~\cite{rf:Skyrme61,rf:Adkins83,rf:Adkins84}.
Following their work, an experimental discovery of the lightest state of $\lbar{\bf 10}$, 
namely $\Theta^{+}(1540)$, was reported by the LEPS collaboration~\cite{rf:Nakano03}.
$\Theta^{+}$ has strangeness $S = +1$ and should contain at least one $\bar{s}$ quark.
It is called an exotic baryon or a pentaquark,
because the minimal number of the quarks is five from the charge and the strangeness. 
Although later many experiments confirmed this finding,
several experiments did not observe $\Theta^{+}$.
Lists of these published experiments and detailed discussion of their results are
presented in Refs.~\cite{rf:Hicks05,rf:Danilov05}. 

Theoretically, there are many works based on
the Skyrme model~\cite{rf:Praszalowicz03},
the diquark models~\cite{rf:diquark},
the chiral bag model~\cite{rf:Hosaka03},
the MIT bag model~\cite{rf:Carlson03},
the constituent quark model~\cite{rf:Stancu03},
the QCD sum rules~\cite{rf:QCDsum},
and the lattice QCD~\cite{rf:laticeQCD}.
These works are reviewed in Ref.~\cite{rf:Oka04}.

We are interested in the descriptions of the $\lbar{\bf 10}$ baryons
by the soliton~\cite{rf:Praszalowicz03,rf:Itzhaki04,rf:Cohen04,rf:Walliser05}
in the SU(3) Skyrme model~\cite{rf:Guadanini84,rf:Mazur84,rf:Manohar84,rf:Chemtob85,rf:Jain85}.
Now, there are two major approaches to quantize the soliton.
First is the Callan-Klebanov approach~\cite{rf:Callan85},
in which baryons appear as kaon-SU(2) Skyrmion bound states,
and the isospin rotation of the soliton and the fluctuations of the kaon field are quantized.
The bound states change according to the baryon states.
In particular the Wess-Zumino term acts as a repulsive force on the $S = +1$ states;
its strength is strong enough to remove all bound states~\cite{rf:Callan88} and
all resonances~\cite{rf:Scoccola90} for the standard values of the parameters.
However, recently, Itzhaki {\it et al.}~\cite{rf:Itzhaki04} applied this approach to the exotic baryons and
found the kaon bound states of $S = +1$ by using a large kaon mass~($\sim 1$~GeV).

Second is the rigid rotator approach~(RRA)~\cite{rf:Adkins83},
in which the shape of the soliton is common to all baryon states
and the rotation of the soliton in flavor space is quantized.
Then, the baryons emerge as the rotational states of a rigid soliton.
From early papers~\cite{rf:Manohar84,rf:Chemtob85} on the SU(3) Skyrme model,
it was pointed out that this approach reproduces not only the octet (${\bf 8}$) and decuplet (${\bf 10}$) baryons
but also the antidecuplet ($\lbar{\bf 10}$) baryons as the low lying spectrum.
The $\lbar{\bf 10}$ baryons have the spin and the parity $J^{P} = 1/2^{+}$ in this approach.

However, a limit of the applicability of RRA has been pointed
out~\cite{rf:Bander84,rf:Braaten85,rf:Rajaraman86} in the SU(2) Skyrme model.
The shape of the soliton changes because of the centrifugal force of the rotation, and
the large spin of the baryon leads to the instability of the soliton
due to the spontaneous emission of the real pion from the soliton.
In the SU(3) Skyrme model, the low lying multiplets (${\bf 8}$, ${\bf 10}$, and $\lbar{\bf 10}$)
seem free from the limit of the applicability of RRA~\cite{rf:Diakonov97} due to their small spins.
However, the situation depends on the baryon states.
The rotation emerges in the strangeness direction simultaneously
and pushes the shape of the soliton out further.
In addition, if the shape of the soliton is affected by the strangeness degrees of freedom,
it would shrink because of the large meson mass.
Therefore, there is a possibility too that the flavor symmetry breaking cancels
the deformation caused by the rotation.
We consider that this possibility should be investigated particularly.

Furthermore, Itzhaki {\it et al.}~\cite{rf:Itzhaki04} and Cohen~\cite{rf:Cohen04} pointed out that
in a large number of the color ($N_{c}$) expansion,
the mass differences between the $\lbar{\bf 10}$ and ${\bf 8}$ baryons scale as $N_{c}^{0}$.
This means that RRA for multiplet $\lbar{\bf 10}$ is not consistent with the large $N_{c}$ expansion.
Since the above mentioned deformations of the classical soliton are
formally sub-leading effects in the expansion,
we are interested in whether the effects could be practically negligible in $\lbar{\bf 10}$.

In this paper, we formulate a mean field approach 
to include the effects of the rotation and the symmetry breaking into the shape of the soliton.
In this meaning, we modify RRA.
In addition, we study numerically the soliton solutions derived from our approach, and
find the input parameters that keep the soliton stable and reproduce the baryon mass spectrum. 

In Sec.~\ref{sec:model}, the SU(3) Skyrme model and its collective quantization are reviewed, and
the mean field approach for the soliton is introduced.
In Sec.~\ref{sec:soliton}, the stability conditions of the soliton solution are explained, and
the numerical solutions are displayed.
In Sec.~\ref{sec:mass}, the input parameters and the resultant baryon mass spectrum are given.
Finally, in Sec.~\ref{sec:summary} we summarize the results.

\section{SU(3) Skyrme model\label{sec:model}}
\subsection{Model and the collective coordinate quantization\label{subsec:quantization}}
The effective action \cite{rf:Praszalowicz85} we take here is given by
\begin{eqnarray}
\Gamma &=& \Gamma_{S} + \Gamma_{SB} + \Gamma_{WZ},
\label{eq:action}\\
\Gamma_{S} &=& \int d^{4}x \left(
\frac{f_{\pi}^{2}}{4} \trace \partial_{\mu} U \partial^{\mu} U^{\dagger}
\right.
\nonumber \\ &&
\left.
+ \frac{1}{32 e^{2}} \trace \comm{U^{\dagger} \partial_{\mu} U}{U^{\dagger} \partial_{\nu} U}^{2}
\right),\\
\Gamma_{SB} &=& \int d^{4}x \left\{
\frac{f_{\pi}^{2}}{8} (m_{\pi}^{2} + m_{\eta}^{2}) \trace (U + U^{\dagger} - 2)
\right.
\nonumber \\ &&
\left.
+\frac{f_{\pi}^{2}}{2 \sqrt{3}} (m_{\pi}^{2} - m_{K}^{2}) \trace \lambda_{8} (U + U^{\dagger})
\right\},\\
\Gamma_{WZ} &=& \frac{i N_{c}}{240 \pi^{2}} \int_{D_{5}} \trace \left(dUU^{\dagger}\right)^{5},
\end{eqnarray}
where $f_{\pi}$ is the pion decay constant, $e$ is the Skyrme parameter,
$U(x)$ is the SU(3) unitary matrix representing the pseudoscalar mesons ($\pi$, $K$, $\eta$),
$m_{\pi,K,\eta}$ are the masses of ($\pi$, $K$, $\eta$),
$\lambda_{8}$ is the 8th component of the Gell-Mann matrices $\lambda_{\mu}$~($\mu = 1, 2, \dots,8$),
and $N_{c}$ is the number of color degrees of freedom.
The Wess-Zumino term $\Gamma_{WZ}$ is given as an integral over the five-dimensional disk $D_{5}$,
the boundary of which is the compactified space-time $S_{4}$.
The symmetry breaking mass term $\Gamma_{SB}$ contains only two masses $m_{\pi,K}$
because of the quadratic sum rule:
$m_{\pi}^{2} + 3 m_{\eta}^{2} - 4 m_{K}^{2} = 0$.
In this paper, we choose $(e, f_{\pi}, m_{\pi}, m_{K})$ as adjustable parameters. 

The effective action~(\ref{eq:action}) admits a classical static soliton solution
under the hedgehog ansatz embedded in the SU(2) subgroup:
\begin{equation}
U(x) \rightarrow U_{0}({\bf r}) = \exp \left[ i \sum_{i=1}^{3} \lambda_{i} \hat{x}_{i} F(r) \right],
\label{eq:u0}
\end{equation}
where $r = |{\bf r}|$, $\hat{x}_{i} = x_{i}/r$,
and $F(r)$ is the profile function of the soliton.
The baryon number one solution is subjected to the boundary conditions
\begin{equation}
F(0) = \pi,~~F(\infty) = 0.
\label{eq:boundary}
\end{equation}

We postulate the cranking form \cite{rf:Adkins83} of the time dependent meson field:
\begin{equation}
U(x) = A(t) U_{0}({\bf r}) A^{\dagger}(t),
\label{eq:cranking}
\end{equation}
where $A(t)$ describes the adiabatic collective rotation of the system in SU(3) flavor space.
Using the standard method \cite{rf:Manohar84,rf:Chemtob85,rf:Guadanini84,rf:Mazur84,rf:Jain85}
to quantize the motion on the SU(3) group manifold,
we obtain a dimensionless quantized collective Hamiltonian $\tilde{H}$
and a first class constraint on the 8th generator $R_{8}$ of ${\rm SU_{R}(3)}$:
\begin{eqnarray}
\tilde{H} &=& \tilde{M}_{0}
+ \frac{e^{4}}{2} \left(
  \frac{1}{\tilde{\alpha}^{2}}-\frac{1}{\tilde{\beta}^{2}}
\right) C_{2}({\rm SU_{R}}(2))
-\frac{e^{4}}{2 \tilde{\beta}^{2}} R_{8}^{2}
\nonumber \\ &&
+ \frac{e^{4}}{2 \tilde{\beta}^{2}} C_{2}({\rm SU_{R}}(3))
+ \frac{\tilde{\gamma}}{2} \left[1-D^{(8)}_{88}(A)\right],
\label{eq:hcoll}\\
R_{8} &=& -\frac{N_{c}}{2 \sqrt{3}},
\label{eq:constraint}
\end{eqnarray}
where $C_{2}({\rm SU_{R}}(3))$ and $C_{2}({\rm SU_{R}}(2))$ are the Casimir operators
of ${\rm SU_{R}}(3)$ and ${\rm SU_{R}}(2)$ respectively.
In addition,
\begin{equation}
D^{(8)}_{88}(A) = \frac{1}{2} \trace \left(\lambda_{8}A^{\dagger}\lambda_{8}A\right),
\end{equation}
and
\begin{eqnarray}
\tilde{M}_{0} &=& 4\pi \int d\rho \rho^{2} \left[
\frac{1}{2} \left(1+2\frac{\sin^{2} F}{\rho^{2}}\right) F'^{2}
\right.
\nonumber \\ &&
\left.
+ \frac{\sin^{2} F}{\rho^{2}} \left(1+\frac{\sin^2 F}{2\rho^{2}}\right)
+ \tilde{m}_{\pi}^{2} (1-\cos F)
\right],
\label{eq:mcl}\\
\tilde{\alpha}^{2} &=& \frac{8 \pi}{3} \int d\rho \rho^{2}
\sin^{2} F \left(F'^{2} + 1 + \frac{\sin^{2} F}{\rho^{2}}\right),
\label{eq:alpha2}\\
\tilde{\beta}^{2} &=& 4\pi \int d\rho \rho^2
\sin^{2} \frac{F}{2} \left(\frac{F'^{2}}{4} + 1 + \frac{\sin^{2} F}{2\rho^{2}}\right),
\label{eq:beta2}\\
\tilde{\gamma} &=& \frac{16 \pi}{3} (\tilde{m}_{K}^{2}-\tilde{m}_{\pi}^{2})
\int d\rho \rho^{2} (1-\cos F),
\label{eq:gamma}
\end{eqnarray}
where $\rho = e f_{\pi} r$, $\tilde{m}_{\pi,K} = m_{\pi,K}/(e f_{\pi})$,
and $F' = \frac{dF}{d\rho}$.
The Hamiltonian $\tilde{H}$ explicitly depends only on $e$ and $\tilde{m}_{\pi,K}$.
The Hamiltonian, the classical soliton mass,
and the symmetry breaking with the physical unit (MeV) are given by
$H = \frac{f_{\pi}}{e} \tilde{H}$,
$M_{0} = \frac{f_{\pi}}{e} \tilde{M}_{0}$,
and
$\gamma = \frac{f_{\pi}}{e} \tilde{\gamma}$
respectively.
The moments of inertia with the physical unit (1/MeV) are given by
$\alpha^{2} = \tilde{\alpha}^{2}/(e^{3} f_{\pi})$
and
$\beta^{2} = \tilde{\beta}^{2}/(e^{3} f_{\pi})$.

The state function of the baryon $B$ is labeled as
\begin{equation}
\Psi_{B} = \Psi\left(
\begin{array}{cc}
IY & JY_{R}\\
I_{3}&J_{3}
\end{array}
\right),
\label{eq:eigenvec}
\end{equation}
where $(J,J_{3})$, $(I,I_{3})$, $Y$, and $Y_{R}$ are the eigenvalues
of the spin, the isospin, the hypercharge, and the right hypercharge, respectively.
The right hypercharge is given by $Y_{R} = 1$ due to constraint~(\ref{eq:constraint}).
Equation~(\ref{eq:eigenvec}) is obtained by solving the following eigenvalue equation,
\begin{equation}
\tilde{H} \Psi_{B} = \tilde{E}_{B} \Psi_{B},
\label{eq:eigeneq}
\end{equation}
where $\tilde{E}_{B}$ is the dimensionless energy eigenvalue of the baryon state $\Psi_{B}$.
To solve this equation, we use the Yabu and Ando method \cite{rf:Yabu88},
in which $\tilde{E}_{B}$ is given by
\begin{eqnarray}
\tilde{E}_{B} &=& \tilde{M}_{0}
+ \frac{e^4}{2} \left(\frac{1}{\tilde{\alpha}^2}-\frac{1}{\tilde{\beta}^2}\right) J (J+1)
\nonumber \\ &&
-\frac{3e^4}{8\tilde{\beta}^{2}} + \frac{e^4}{2\tilde{\beta}^{2}} {\cal E}_{SB}.
\label{eq:EB}
\end{eqnarray}
Quantity ${\cal E}_{SB}$ is the dimensionless eigenvalue of
\begin{equation}
	\left[C_{2}({\rm SU_{R}}(3)) + \frac{\tilde{\beta}^{2} \tilde{\gamma}}{e^4} (1-D^{(8)}_{88}(A))\right]\Psi_{B}
	= {\cal E}_{SB} \Psi_{B}.
\label{eq:YA}
\end{equation}

\subsection{Mean field approach to the baryon states\label{subsec:meanfield}}
To solve the eigenvalue Eq.~(\ref{eq:eigeneq}) and 
obtain the baryon states~\cite{rf:Adkins83,rf:Adkins84,rf:Yabu88},
one should know about the profile function $F(\rho)$ in Eqs.~(\ref{eq:mcl})-(\ref{eq:gamma}).
Here, we define an equation of motion for $F(\rho)$ as
\begin{equation}
\frac{\delta \tilde{{\cal H}}_{B}}{\delta F(\rho)} = 0,
\label{eq:eom}
\end{equation}
where $\tilde{{\cal H}}_{B}$ is a {\it classical} Hamiltonian for each baryon $B$.
In RRA~\cite{rf:Adkins83}, $\tilde{{\cal H}}_{B} = \tilde{M}_{0}$, and
$F$ is not affected by the rotation and the symmetry breaking at all.
Therefore, $F$ is common to all baryons.

In this paper, we adopt the following mean field Hamiltonian~\cite{rf:Akiyama03,rf:Akiyama04,rf:Akiyama06},
\begin{equation}
\tilde{{\cal H}}_{B} = \braketo{\Psi^{(0)}_{B}}{\tilde{H}}{\Psi^{(0)}_{B}},
\label{eq:hb_def}
\end{equation}
where $\ket{\Psi^{(0)}_{B}}$ is an eigenstate of $\tilde{H}$
without the SU(3) symmetry breaking ($\tilde{\gamma} = 0$), and
the state is represented by the SU(3) $D$ function corresponding to the baryon $B$.

Our classical Hamiltonian $\tilde{{\cal H}}_{B}$ includes the influence of the flavor symmetry breaking
of the first order in powers of $ (\tilde{m}_{K}^{2}-\tilde{m}_{\pi}^{2})$
and the influence of the rotation of order $1/N_{c}$ in the large $N_{c}$ expansion.
The specific expression of $\tilde{{\cal H}}_{B}$ is given by
\begin{equation}
\tilde{{\cal H}}_{B} = \tilde{M}_{0}
+ \frac{e^{4}}{2} \left(
	\tilde{\alpha}^{2} \tilde{\omega}^{2}
	+ \tilde{\beta}^{2} \tilde{\kappa}^{2}
\right)
+ \frac{3}{4} q_{B} \tilde{\gamma},
\label{eq:hb}
\end{equation}
where
\begin{eqnarray}
\tilde{\omega}^{2} &=& \frac{1}{\tilde{\alpha}^{4}} J (J+1),
\label{eq:omega}\\
\tilde{\kappa}^{2} &=& \frac{1}{\tilde{\beta}^{4}} \left[
	C_{2}({\rm SU_{R}}(3))-J (J+1)-\frac{3}{4}
\right],
\label{eq:kappa}\\
q_{B} &=& \frac{2}{3} \braketo{\Psi^{(0)}_{B}}{1-D^{(8)}_{88}(A)}{\Psi^{(0)}_{B}}.
\label{eq:qparam}
\end{eqnarray}

The quantities $\tilde{\omega}$ and $\tilde{\kappa}$ distinguish
the multiplets (${\bf 8}$, ${\bf 10}$, $\lbar{\bf 10}$), and
the values of $J (J+1)$ and $C_{2}({\rm SU_{R}}(3))-J (J+1)-\frac{3}{4}$ are given at Table~\ref{tb:wkparam}.
For the states with $J = I$, $\tilde{\omega}$ and $\tilde{\kappa}$ are regarded as
the angular frequencies of the rotation in ordinary space and strangeness direction, respectively.
The expectation value $q_{B}$ is a source of the SU(3) symmetry breaking on the profile function
and characterizes each baryon state.
Table~\ref{tb:qparam} shows the values of $q_{B}$ for the individual ${\bf 8}$, ${\bf 10}$, and $\lbar{\bf 10}$ baryons.
Therefore, the profile functions derived from the classical Hamiltonian~(\ref{eq:hb}) change the shapes
according to the baryon states.
Our approach modifies RRA in this meaning.

\begin{table}
\caption{SU(3) representation $(p,q)$, spin $J$,
$J (J+1)$ in Eq.~(\ref{eq:omega}),
and $C_{2}({\rm SU_{R}}(3))-J (J+1)-\frac{3}{4}$ in Eq.~(\ref{eq:kappa})
for multiplets ${\bf 8}$, ${\bf 10}$, and $\lbar{\bf 10}$.}
\begin{ruledtabular}
\begin{tabular}{ccccc}
& $(p,q)$ & $J$ & $J (J+1)$ & $C_{2}({\rm SU_{R}}(3))-J (J+1)-\frac{3}{4}$ \\
\hline
${\bf  8}$ & $(1,1)$ & $1/2$ & $3/4$ & $3/2$\\
${\bf 10}$ & $(3,0)$ & $3/2$ & $15/4$ & $3/2$\\
$\lbar{\bf 10}$ & $(0,3)$ & $1/2$ & $3/4$ & $9/2$
\end{tabular}
\end{ruledtabular}
\label{tb:wkparam}
\end{table}   

\begin{table}
\caption{Expectation value $q_{B}$ of Eq.~(\ref{eq:qparam}),
where $B \in {\bf 8}, {\bf 10}, \lbar{\bf 10}$.}
\begin{ruledtabular}
\begin{tabular}{ccccc}
$B \in {\bf  8}$ & $N$ & $\Lambda$ & $\Sigma$ & $\Xi$ \\
$q_{B}$ & $7/15$ & $9/15$ & $11/15$ & $12/15$\\
\hline
$B \in {\bf 10}$ & $\Delta$ & $\Sigma^{*}$ & $\Xi^{*}$ & $\Omega$ \\
$q_{B}$ & $7/12$ & $8/12$ & $9/12$ & $10/12$\\
\hline
$B \in \lbar{\bf 10}$ & $\Theta^{+}$ & $N^{*}_{\lbar{10}}$ & $\Sigma^{*}_{\lbar{10}}$ & $\Xi^{*}_{\lbar{10}}$ \\
$q_{B}$ & $6/12$ & $7/12$ & $8/12$ & $9/12$
\end{tabular}
\end{ruledtabular}
\label{tb:qparam}
\end{table}   

We obtain the equation of motion for $F(\rho)$ from Eqs.~(\ref{eq:eom}) and (\ref{eq:hb}):
\begin{equation}
C_{F''}(\rho, F) F'' + C_{\left(F'\right)^{2}}(\rho, F) \left(F'\right)^{2} + C_{F'}(\rho, F) F' + C(\rho, F) = 0,
\label{eq:eomF}
\end{equation}
where
\begin{eqnarray}
C_{F''}(\rho, F) &=&
	1 + \frac{2 \sin^{2} F}{\rho^{2}}
	-e^{4} \left(
		\frac{2}{3} \tilde{\omega}^{2} \sin^{2} F
		+\frac{1}{4} \tilde{\kappa}^{2} \sin^{2} \frac{F}{2}
	\right),\\
C_{\left(F'\right)^{2}}(\rho, F) &=&
\frac{\sin 2 F}{\rho^{2}}
	-e^{4} \left(
		\frac{1}{3} \tilde{\omega}^{2} \sin 2F
		+\frac{1}{16} \tilde{\kappa}^{2} \sin F
	\right),\\
C_{F'}(\rho, F) &=&
\left[
	1-e^{4} \left(
		\frac{2}{3} \tilde{\omega}^{2} \sin^{2} F
		+\frac{1}{4} \tilde{\kappa}^{2} \sin^{2} \frac{F}{2}
	\right)
\right] \frac{2}{\rho},\\
C(\rho, F) &=&
-\left(1 + \frac{\sin^{2} F}{\rho^{2}}\right) \frac{\sin 2F}{\rho^2}
-\tilde{m}_{\rm eff}^{2} \sin F
\nonumber \\ &&
+\frac{1}{3} e^{4} \tilde{\omega}^{2}
	\left(1 + \frac{2 \sin^{2} F}{\rho^{2}}\right) \sin 2F
\nonumber \\ &&
+\frac{1}{4} e^{4} \tilde{\kappa}^{2} \left[
	1 + \left(1 + 3 \cos F\right) \frac{\sin^{2} \frac{F}{2}}{\rho^{2}} 
\right] \sin F,
\end{eqnarray}
and $\tilde{m}_{\rm eff}$ is an effective meson mass given by
\begin{equation}
\tilde{m}_{\rm eff}^{2} = \tilde{m}_{\pi}^{2} (1-q_{B}) + \tilde{m}_{K}^{2} q_{B}.
\label{eq:meff}
\end{equation}

Coupled equations (\ref{eq:alpha2}), (\ref{eq:beta2}), (\ref{eq:omega}), (\ref{eq:kappa}), and (\ref{eq:eomF})
are self-consistently solved under the boundary conditions~(\ref{eq:boundary}).
Then, independent parameters are $(e, \tilde{m}_{\rm eff})$ only.
Therefore the effect of the flavor symmetry breaking on the profile function is expressed by $\tilde{m}_{\rm eff}$.
In RRA, $\tilde{m}_{\rm eff} = m_{\rm eff}/(e f_{\pi}) = m_{\pi}/(e f_{\pi}) \ll 1$.
In our approach, the effective mass can take a value $\sim \tilde{m}_{K}$ according to $q_{B}$.
This fact is important.
If we estimate that $e = 3.87$, $f_{\pi} = 44.5$~MeV, and $m_{K} = 495$~MeV~\cite{rf:Yabu88},
a large effective mass $\tilde{m}_{\rm eff} \sim 2.8$ is obtained.
Therefore, we can expect qualitatively different behavior of the soliton solution in our approach.

\section{Soliton solution in the mean field approach\label{sec:soliton}}
Our next task is to perform the self-consistent procedure in Sec.~\ref{subsec:meanfield}.
The procedure is faced with two kinds of complexity:
the instability of the soliton solution and
the dependence of the soliton solution on the multiplets.
The instability results from Eqs.~(\ref{eq:eomF}) and (\ref{eq:boundary}).
The dependence on the multiplet is brought into the calculation
by Eqs.~(\ref{eq:omega}) and (\ref{eq:kappa}).
Therefore, we discuss these problems separately in the following sections.

\subsection{Instability of the soliton solution\label{subsec:instability}}
To investigate the instability of the soliton solution,
we treat $(e^{4} \tilde{\omega}^{2}, e^{4} \tilde{\kappa}^{2}, \tilde{m}_{\rm eff}^{2})$
as input parameters in this subsection and the next one.
The parameter space is designated as ${\cal M}$.

The stable soliton solutions of Eq.~(\ref{eq:eomF}) are obtained only in a restricted area of ${\cal M}$.
The restriction has two origins.
One is a behavior of the profile function $F$ at $\rho \sim \infty$
due to the centrifugal force of the rotation \cite{rf:Bander84,rf:Braaten85,rf:Rajaraman86}:
\begin{eqnarray}
F &\sim& \frac{A}{\rho^{2}} (1+\mu \rho) e^{-\mu \rho},
\label{eq:asymF}\\
\mu &=& \sqrt{\tilde{m}_{\rm eff}^{2}-e^{4} \left(
	\frac{2}{3} \tilde{\omega}^{2}+\frac{1}{4} \tilde{\kappa}^{2}
\right)}.
\end{eqnarray}
Here the rotation pushes $F$ out of the center of the soliton.
For the stable soliton solution, the following condition should be satisfied;
\begin{equation}
\tilde{m}_{\rm eff}^{2}-e^{4} \left(
	\frac{2}{3} \tilde{\omega}^{2}+\frac{1}{4} \tilde{\kappa}^{2}
\right) \geq 0.
\label{eq:stcond1}
\end{equation}
Therefore, the rotating SU(3) Skyrmion with $\tilde{m}_{\rm eff} = 0$ is unstable.
That is analogous to the result of the rotating SU(2) Skyrmion
in the chiral limit \cite{rf:Bander84}.
The rotating SU(3) Skyrmion, however,
can exist in a limit $(\tilde{m}_{\pi} = 0, \tilde{m}_{K} \not= 0)$,
because $\tilde{m}_{\rm eff}^{2} > 0$ from Eq.(\ref{eq:meff}) and Table~\ref{tb:qparam}.

Another origin of the restriction on ${\cal M}$ is
a behavior of the coefficient function of $F''$ in Eq.~(\ref{eq:eomF}). 
We have the second condition
for the stable soliton solution satisfying the boundary conditions~(\ref{eq:boundary}):
\begin{equation}
C_{F''}(\rho, F(\rho)) > 0.
\label{eq:stcond2}
\end{equation}
For verification of this condition, we define a curve $F^{*}$ in $(\rho, F)$ plane by
\begin{equation}
C_{F''}(\rho, F^{*}) = 0.
\label{eq:cofddF0}
\end{equation}
Function $F^{*}$ should not be confused with the profile function $F$.
We will show that $F$ satisfying the boundary conditions~(\ref{eq:boundary}) cannot cross $F^{*}$
and this requirement is equal to condition~(\ref{eq:stcond2}).

\begin{figure}
	\includegraphics[clip]{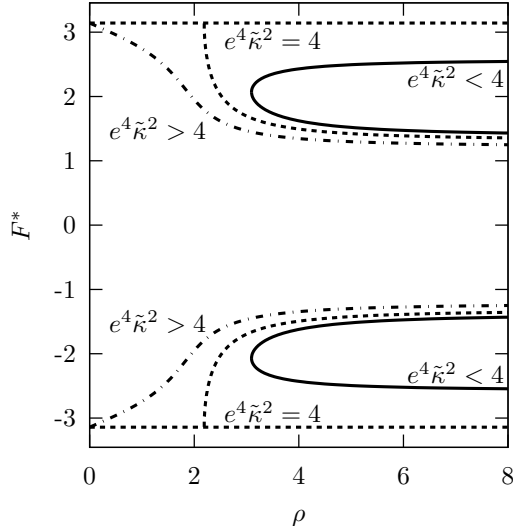}
	\caption{Typical forms of $F^{*}$ within $-\pi \leq F^{*} \leq \pi$.
	The cases of $e^{4} \tilde{\kappa}^{2} < 4$,
	$e^{4} \tilde{\kappa}^{2} = 4$, and $e^{4} \tilde{\kappa}^{2} > 4$ are
	represented by the solid lines, the dashed lines, and the dash-dotted lines, respectively.
	For classification of the lines, we set $e^4 \tilde{\omega}^2 \sim 1$.
	}
\label{fig:s_curve_b_dep}
\end{figure}

\begin{figure}
	\includegraphics[clip]{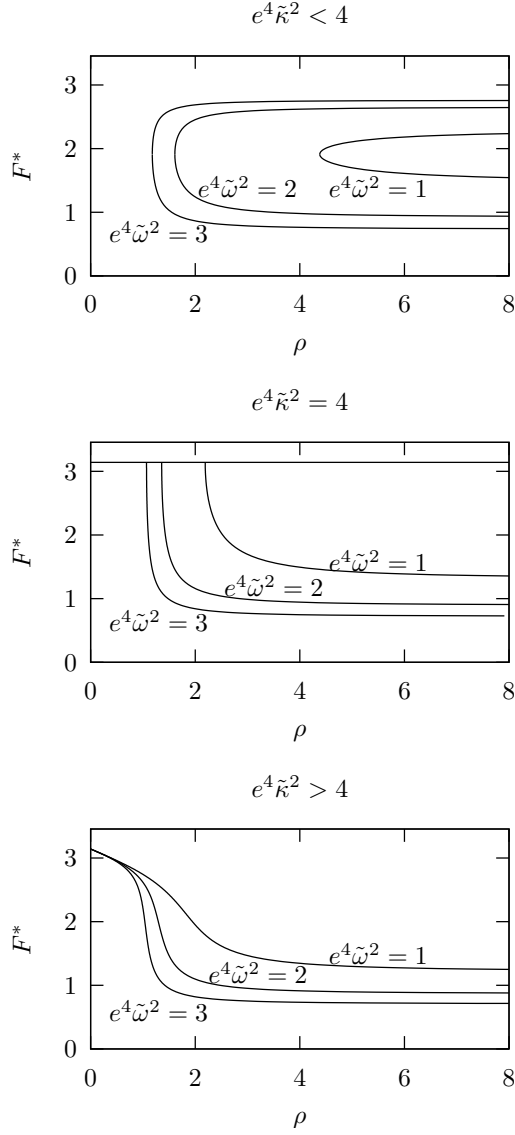}
	\caption{Typical dependence of $F^{*}$ on $e^4 \tilde{\omega}^2$ in the range $0 \leq F^{*} \leq \pi$
	for $e^{4} \tilde{\kappa}^{2} < 4$, $e^{4} \tilde{\kappa}^{2} = 4$, and $e^{4} \tilde{\kappa}^{2} > 4$.
	Actual values of $e^{4} \tilde{\kappa}^{2}$ in the figures are 3, 4, 5, respectively.
	}
\label{fig:s_curve_a_dep}
\end{figure}

At first, we investigate properties of $F^{*}$.
Since Eq.~(\ref{eq:cofddF0}) is a quadratic equation for $\sin^{2} \frac{F^{*}}{2}$,
it has two formal solutions for fixed  $\rho$ and $(e^{4} \tilde{\omega}^{2}, e^{4} \tilde{\kappa}^{2})$.
These formal solutions, however, do not always support two real number values
of $F^{*}$ in the range $0 \leq F^{*} \leq \pi$
which has one-to-one correspondence with the range
$0 \leq \sin^{2} \frac{F^{*}}{2} \leq 1$.
If $F^{*}$ is a real number solution in the range,
$\pm F^{*} + 2n\pi$ ($n$: integer) also are solutions in other range.
Practically, we can restrict the value of $F^{*}$ to the range $-\pi \leq F^{*} \leq \pi$,
because the boundary conditions~(\ref{eq:boundary}) ensure that the value of $F$ is in the range.
Figure~\ref{fig:s_curve_b_dep} shows the typical forms of $F^{*}$
for $e^{4} \tilde{\kappa}^{2} < 4$, $e^{4} \tilde{\kappa}^{2} = 4$, and $e^{4} \tilde{\kappa}^{2} > 4$.
The form of $F^{*}$ changes drastically at $e^4 \tilde{\kappa}^2 = 4$.
In particular, there are the constant solutions $F^{*} = \pm \pi$ for $e^{4} \tilde{\kappa}^{2} = 4$,
and $F^{*}$ for $e^4 \tilde{\kappa}^2 \geq 4$ always reaches $\rho = 0$.
Figure~\ref{fig:s_curve_a_dep} show also the dependence of $F^{*}$ on $e^{4} \tilde{\omega}^{2}$
in the range $0 \leq F^{*} \leq \pi$.
For a larger value of $e^{4} \tilde{\omega}^{2}$, $F^{*}$  becomes
closer to the axes $\rho = 0$ and $F^{*} = 0$.

Next, we explain how to verify Eq.~(\ref{eq:stcond2}).
If $F$ crosses $F^{*}$ at some radius $\rho = \rho^{*}$,
Eq.~(\ref{eq:eomF}) becomes a quadratic equation for $F'(\rho^{*})$:
\begin{equation}
C_{\left(F'\right)^{2}}(\rho^{*}, F^{*}) \left(F'\right)^{2} + C_{F'}(\rho^{*}, F^{*}) F' + C(\rho^{*}, F^{*}) = 0.
\label{eq:eomF0}
\end{equation}
Then we can statically calculate the value of the discriminant
\begin{equation}
D^{*} = C_{F'}(\rho^{*}, F^{*})^{2} - 4 C_{\left(F'\right)^{2}}(\rho^{*}, F^{*}) C(\rho^{*}, F^{*}).
\label{eq:dis0}
\end{equation}
Of course, $D^{*} < 0$ means that Eq.~(\ref{eq:eomF0}) has no real number solution and
$F$ cannot cross  $F^{*}$ at $\rho = \rho^{*}$ from the beginning.
Moreover Eq.~(\ref{eq:eomF}) does not have the real number solution in a neighborhood of the point,
because its discriminant for $F'$:
\begin{eqnarray}
C_{F'}(\rho, F)^{2} &-& 4 C_{\left(F'\right)^{2}}(\rho, F) C(\rho, F)
\nonumber \\
&-& 4 C_{\left(F'\right)^{2}}(\rho, F) C_{F''}(\rho, F) F''
\label{eq:dis}
\end{eqnarray}
approaches the value of $D^{*}~(< 0)$ near $F^{*}$,
and the value of $F'(\rho)$ becomes complex numbers.

For $D^{*} \geq 0$, the values of $F'(\rho^{*})$ are real numbers.
However it is analytically unclear whether the values are consistent
with the boundary conditions~(\ref{eq:boundary}).
From numerical calculations, we conclude also here that $F$ and $F^{*}$ cannot cross.

\begin{figure}
	\includegraphics[clip]{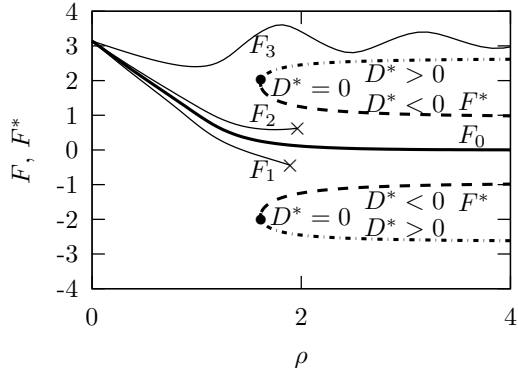}
	\caption{
	Profile functions $F_{i}$~($i = 0,1,2,3$) and curves $F^{*}$ for $e^4 \tilde{\kappa}^2 < 4$
	($e^4 \tilde{\kappa}^2 = 3$, $e^4 \tilde{\omega}^2 = 2$, $\tilde{m}_{\rm eff} = 2$).
	The solid lines represent $F_{i}$.
	The dashed lines indicate $D^{*} < 0$ parts of $F^{*}$,
	the dash-dotted lines $D^{*} > 0$ parts, and
	marks ``$\bullet$'' represent the points at which $D^{*} = 0$.
	Marks ``$\times$'' represent breakdowns of Eq.~(\ref{eq:eomF}) at these points.
	}
\label{fig:scattering_b_lt_4}
\end{figure}

\begin{figure}
	\includegraphics[clip]{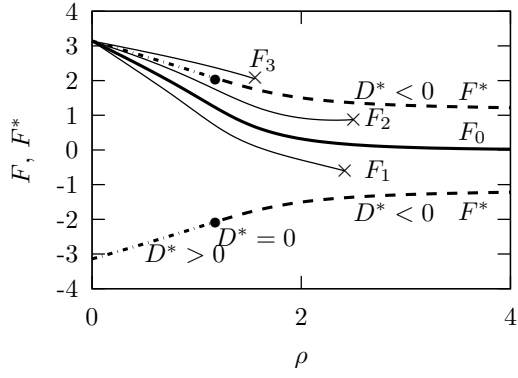}
	\caption{
	Profile functions $F_{i}$~($i = 0,1,2,3$) and curves $F^{*}$ for $e^4 \tilde{\kappa}^2 > 4$
	($e^4 \tilde{\kappa}^2 = 10$, $e^4 \tilde{\omega}^2 = 0.5$, $\tilde{m}_{\rm eff} = 1.9$).
	Meanings of the figure symbols are the same as those of Fig.~\ref{fig:scattering_b_lt_4}.
	}
\label{fig:scattering_b_gt_4}
\end{figure}

Figures~\ref{fig:scattering_b_lt_4} and \ref{fig:scattering_b_gt_4} show the examples.
Profile functions $F_{i}$~($i = 0,1,2,3$) and $F^{*}$ for $e^4 \tilde{\kappa}^2 < 4$ are plotted
in Fig.~\ref{fig:scattering_b_lt_4}.
Each of $F_{i}$ corresponds to different values of $F'(0)$, namely
\begin{equation}
0 > F'_{3}(0) > F'_{2}(0)  > F'_{0}(0)  > F'_{1}(0) .
\label{eq:df0}
\end{equation}
Only $F_{0}$ satisfies the boundary conditions~(\ref{eq:boundary}).
In the cases of $F_{1,2}$, there are breakdowns of Eq.~(\ref{eq:eomF}),
because the profile function approaches the $D^{*} < 0$ part of $F^{*}$.
Profile function $F_{3}$ survives but does not satisfy the boundary conditions~(\ref{eq:boundary})
due to the scattering by the $D^{*} > 0$ part.

Figure~\ref{fig:scattering_b_gt_4} shows $F_{i}$~($i = 0,1,2,3$)
and $F^{*}$ for $e^4 \tilde{\kappa}^2 > 4$.
Also here, only $F_{0}$ satisfies the boundary conditions~(\ref{eq:boundary}), and
$F_{1,2}$ break because of the same reason as that of the case $e^4 \tilde{\kappa}^2 < 4$.
Moreover, there is a new situation that $F^{*}$ divides $F_{3}$
from the other profiles (e.g. $F_{2}$).
Therefore, $F_{2}$ is in $C_{F''} > 0$ area and  $F_{3}$ is in $C_{F''} < 0$ area.
Although the values of $F'_{2}(0)$ and $F'_{3}(0)$ are close,
the profile functions $F_{2}$ and $F_{3}$ separate with an increase in $\rho$.
It seems that there is a repulsion between $F$ and  the $D^{*} > 0$ part of $F^{*}$.
Profile function $F_{3}$ survives near the $D^{*} > 0$ part at small $\rho$
but breaks near the $D^{*} < 0$ part at last.

From these results, we conclude that $F$ satisfying the boundary conditions~(\ref{eq:boundary})
cannot cross $F^{*}$.
Therefore, $F^{*}$ divides $(\rho, F)$ plane into two areas:
$C_{F''} > 0$ area and $C_{F''} < 0$ area.
Profile function $F(\rho)$ lives in only one area containing the boundary point
$(\rho, F) = (\infty, 0)$ of Eq.~(\ref{eq:boundary}).
Since $C_{F''}(\infty, 0) = 1 > 0$ at the boundary point,
we obtain condition~(\ref{eq:stcond2}) of the stable soliton solution.

We should choose the value of $F'(0)$ carefully
so that $F$ is away from the $D^{*} < 0$ part  of $F^{*}$.
Since $F^{*}$ approaches the axes $\rho = 0$ and $F^{*} = 0$
at the larger values of $e^4 \tilde{\omega}^2$ and $e^4 \tilde{\kappa}^2$
(Figs.~\ref{fig:s_curve_b_dep} and \ref{fig:s_curve_a_dep}),
the choice of $F'(0)$ becomes more difficult.
This situation improves for larger value of $\tilde{m}_{\rm eff}$, because $F$ dumps faster
according to Eq.~(\ref{eq:asymF}).
Thus, with an increase in $\tilde{m}_{\rm eff}$,
the area given by Eq.~(\ref{eq:stcond2}) enlarges with $\tilde{m}_{\rm eff}$
in parameter space ${\cal M}$.

\subsection{Profile function of the stable soliton solution\label{subsec:stablesoliton}}
\begin{figure}
	\includegraphics[clip]{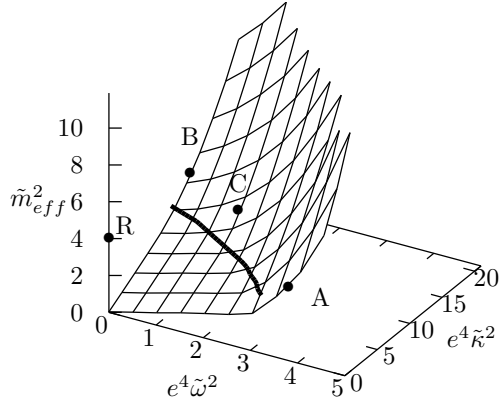}
	\caption{
	Critical surface given by Eqs.~(\ref{eq:eomF}) and (\ref{eq:boundary}) in parameter space ${\cal M}$.
	The sizes of the mesh on the surface are
	$(\Delta e^4 \tilde{\omega}^2, \Delta  e^4 \tilde{\kappa}^2) = (0.5, 2.0)$.
	The area of the stable soliton is on the upper side of the surface.
	The solid line on the critical surface indicates the boundary curve ($\tilde{m}_{\rm eff} \sim 1.6$)
	at which two conditions~(\ref{eq:stcond1}) and (\ref{eq:stcond2}) change the roles.
	Point R is placed at $(0, 0, 4)$.
	Points A, B, and C are placed at $(3.5, 0, 4)$, $(0, 12.7, 4)$, and $(1.5, 9, 4)$
	respectively, and are slightly above the surface.
	}
\label{fig:critical_surface}
\end{figure}

In this subsection, we discuss several constraints
on the parameter space ${\cal M}$ and the stable soliton solutions of Eq.~(\ref{eq:eomF}).
At this stage, we have two conditions~(\ref{eq:stcond1}) and (\ref{eq:stcond2})
for the stable soliton solutions.
Condition~(\ref{eq:stcond1}) is explicitly parametrized
by $(e^4 \tilde{\omega}^2, e^4 \tilde{\kappa}^2, \tilde{m}_{\rm eff}^{2})$, and
condition~(\ref{eq:stcond2}) implicitly.
Both the conditions define together an area of the stable soliton in ${\cal M}$.
In Fig.~\ref{fig:critical_surface}, we show a critical surface
that separates the areas of the stable solution and the unstable solution in ${\cal M}$.
The area of the stable soliton is on the upper side of the surface.

Condition~(\ref{eq:stcond1}) defines the critical surface for small value of $\tilde{m}_{\rm eff}$,
and condition~(\ref{eq:stcond2}) does so for large value of $\tilde{m}_{\rm eff}$.
There is a boundary curve, at which these conditions change the roles on the critical surface.
The boundary has $\tilde{m}_{\rm eff} \sim 1.6$.
We designate the lower (higher) critical surface as ${\cal S}_{L(H)}$.
${\cal S}_{L}$ is a plane, because Eq.~(\ref{eq:stcond1}) is a linear equation
for $(e^4 \tilde{\omega}^2, e^4 \tilde{\kappa}^2, \tilde{m}_{\rm eff}^{2})$.
${\cal S}_{H}$ curves upward steeply.

\begin{figure*}
	\includegraphics[clip]{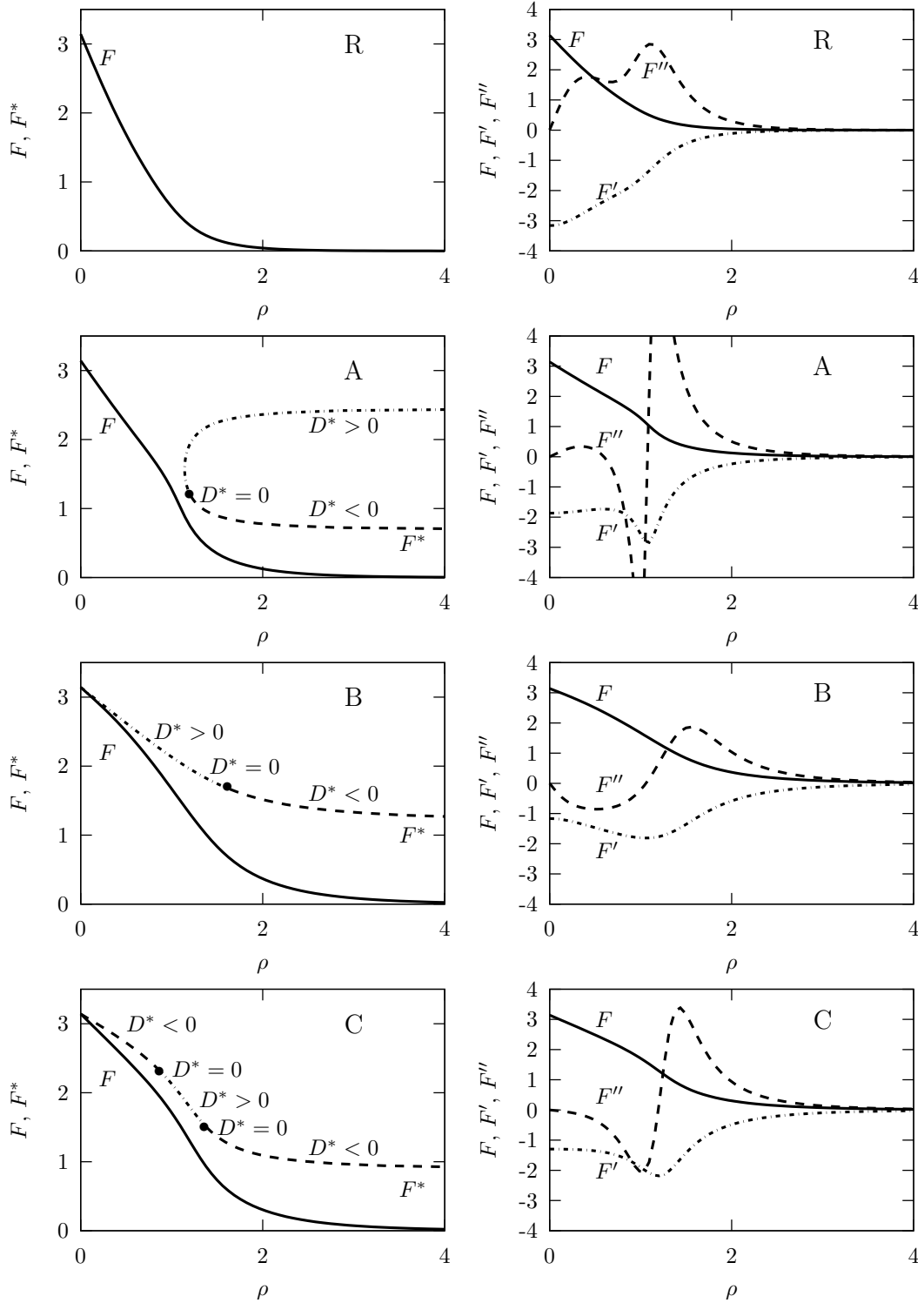}
	\caption{
	Left panel: $F$ and $F^{*}$ for points R, A, B, and C depicted in Fig.~\ref{fig:critical_surface}.
	Right panel: $F$, $F'$, and $F''$ for the points.
	}
\label{fig:near_critical}
\end{figure*}

Here, we illustrate the effects of the rotation and the symmetry breaking
with the four points R, A, B, and C depicted in Fig.~\ref{fig:critical_surface}.
Point R is placed at $(e^4 \tilde{\omega}^2, e^4 \tilde{\kappa}^2, \tilde{m}_{\rm eff}^{2}) = (0, 0, 4)$ in ${\cal M}$.
Points A, B, and C are placed at $(3.5, 0, 4)$, $(0, 12.7, 4)$, and $(1.5, 9, 4)$ respectively,
and they are slightly above the critical surface ${\cal S}_{H}$.
Figure~\ref{fig:near_critical} shows the profile function, its derivatives,
and curve $F^{*}$ [Eq.~(\ref{eq:cofddF0})] with the parameters
corresponding to points R, A, B, and C in $\cal M$.

Case R corresponds to RRA, because the influence of the rotation is ignored.
In Fig.~\ref{fig:near_critical}, there is a flat part of $F''$ at $\rho \sim 0.5$.
It is caused by the large meson mass ($\tilde{m}_{\rm eff} \sim 2$).
Although such a large meson mass is unfamiliar in other studies on the profile function,
it is legitimate in our approach as noted in Sec.~\ref{subsec:meanfield}.
If $\tilde{m}_{\rm eff}$ increases more, the flat part dents downward.
However, $F''$ does not cross the zero, and $F'$ monotonically increases.

The profile function with small $\tilde{m}_{\rm eff}$ (the parameters near ${\cal S}_{L}$) is given
by spreading the solution of case R according to Eq.~(\ref{eq:asymF}).
Then, the effects of the rotation and the symmetry breaking appear
in the single mass parameter $\mu$.
If $\tilde{m}_{\rm eff}$ decreases further, $\mu$ becomes a complex number and
the soliton becomes unstable because of the emission of the real meson.

In the cases of A, B, and C, there are further characteristic behaviors of the profile functions.
Since points A, B, and C are close to ${\cal S}_{H}$ in $\cal M$,
the corresponding profile functions emphasize features of the rotation.

Gradients $|F'(0)|$ in the cases of A, B, and C are small compared with that in the case of R,
because the rotation pushes the profile function out of the center of the soliton.
Then $F''$ should change largely for $F$ to maintain the asymptotic form~(\ref{eq:asymF})
with the large $\tilde{m}_{\rm eff}$.
Therefore, $F''$ crosses the zero, and the behavior of $F'$ becomes complex.

Case A represents the profile functions deformed by the rapid spatial rotation
($e^{4} \tilde{\kappa}^{2} < 4$ and large $e^{4} \tilde{\omega}^{2}$).
Curvature $F''$ change intensely at intermediate $\rho$ region.
This profile function is apparent at this stage, but
it is excluded by the self-consistent procedure as noted in the next subsection.
Moreover, such a profile function is physically unimportant;
point A in Fig.~\ref{fig:critical_surface} is
on $e^{4} \tilde{\kappa}^{2} = 0$ plane corresponding to the SU(2) Skyrmion and
its effective mass is large ($\tilde{m}_{\rm eff} \sim 2$),
however, physically $\tilde{m}_{\rm eff} = m_{\pi}/(e f_{\pi}) \ll 1$ in this sector.

Case B represents the profile functions
affected by the rapid flavor rotation
($e^{4} \tilde{\kappa}^{2} > 4$ and small $e^{4} \tilde{\omega}^{2}$).
Curvature $F''$ is already negative at $\rho \sim 0$.
Since $F^{*}$ reaches $\rho = 0$,
Eq.~(\ref{eq:stcond2}) reduces to
\begin{equation}
1+2 F'(0)^{2}-\frac{e^{4} \tilde{\kappa}^{2}}{4} > 0,
\label{eq:stcond2a}
\end{equation}
according to the boundary conditions~(\ref{eq:boundary}).
If $\tilde{m}_{\rm eff}^{2}$ decreases,
$F'(0)^{2}$ becomes too small and  Eq.~(\ref{eq:stcond2a}) fails.

Case C represents the profile functions affected by both the spatial and the flavor rotation
($e^{4} \tilde{\kappa}^{2} > 4$ and medium value of $e^{4} \tilde{\omega}^{2}$).
Although  $F^{*}$ reaches $\rho = 0$ and Eq.~(\ref{eq:stcond2a}) is valid here too,
the spatial rotation affects the profile function at $\rho \sim 1$ and $F''$ changes greatly there.
This spatial rotation has smaller angular frequency than that of case A.
Therefore, the flavor rotation enhances the effect of the spatial rotation.

\subsection{Self-consistent soliton solution and
the dependence of the classical soliton on the multiplets\label{subsec:multiplet_dep}}
We are ready to study the self-consistent solution of the coupled equations
Eqs.~(\ref{eq:alpha2}), (\ref{eq:beta2}), (\ref{eq:omega}), (\ref{eq:kappa}), (\ref{eq:eomF}),
and the boundary conditions~(\ref{eq:boundary}).
The independent parameters reduce from $(e^4 \tilde{\omega}^2, e^4 \tilde{\kappa}^2, \tilde{m}_{\rm eff}^{2})$
to $(e, \tilde{m}_{\rm eff})$, and
the quantities $(\tilde{\omega}^{2}, \tilde{\kappa}^{2})$ are self-consistently determined
for each multiplet ${\bf 8}$, ${\bf 10}$, and $\lbar{\bf 10}$.
From Table~\ref{tb:wkparam} and Eqs.~(\ref{eq:omega}) and (\ref{eq:kappa}),
one can estimate that
\begin{subequations}
\begin{align}
\left. \tilde{\omega}^{2} \right|_{\bf 10} &\sim 5 \times \left. \tilde{\omega}^{2} \right| _{\bf 8},
\label{eq:w2_8_10}\\
\left. \tilde{\kappa}^{2} \right|_{\bf 10} &\sim \left. \tilde{\kappa}^{2} \right| _{\bf 8}
\label{eq:k2_8_10}
\end{align}
\label{eq:wk2_8_10}
\end{subequations}
for ${\bf 8}$ and ${\bf 10}$, and
\begin{subequations}
\begin{align}
\left. \tilde{\omega}^{2} \right|_{\lbar{\bf 10}} &\sim \left. \tilde{\omega}^{2} \right|_{\bf 8},
\label{eq:w2_8_10bar}\\
\left. \tilde{\kappa}^{2} \right|_{\lbar{\bf 10}} &\sim 3 \times \left. \tilde{\kappa}^{2} \right|_{\bf 8}
\label{eq:k2_8_10bar}
\end{align}
\label{eq:wk2_8_10bar}
\end{subequations}
for ${\bf 8}$ and $\lbar{\bf 10}$, and
\begin{subequations}
\begin{align}
\left. \tilde{\omega}^{2} \right|_{\lbar{\bf 10}} &\sim \frac{1}{5} \times \left. \tilde{\omega}^{2} \right| _{\bf 10},
\label{eq:w2_10_10bar}\\
\left. \tilde{\kappa}^{2} \right|_{\lbar{\bf 10}} &\sim 3 \times \left. \tilde{\kappa}^{2} \right| _{\bf 10}
\label{eq:k2_10_10bar}
\end{align}
\label{eq:wk2_10_10bar}
\end{subequations}
for ${\bf 10}$ and $\lbar{\bf 10}$.

\begin{figure}
	\includegraphics[width=5cm,clip]{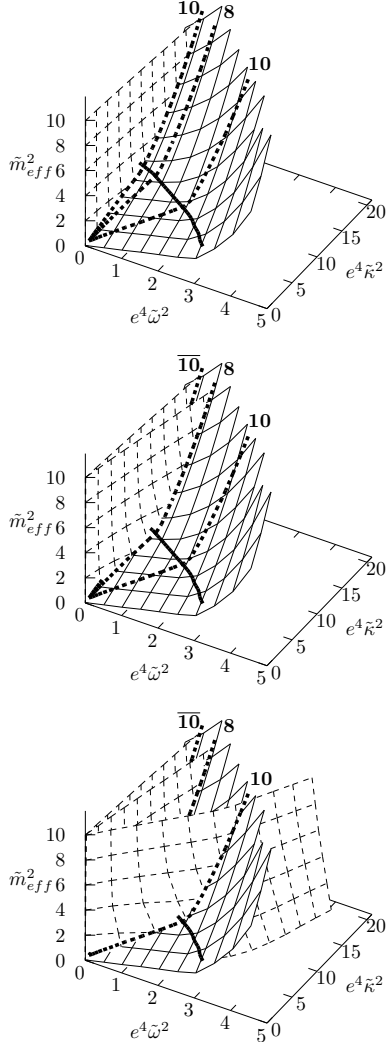}
	\caption{
	Self-consistent surfaces for the multiplets ${\bf 8}$, ${\bf 10}$, and $\lbar{\bf 10}$
	in parameter space ${\cal M}$.
	The dashed lines represent the self-consistent surfaces.
	Three panels show these surfaces one by one in the order of $\lbar{\bf 10}$, ${\bf 8}$, and ${\bf 10}$.
	The solid lines denote the critical surface in Fig.~\ref{fig:critical_surface}.
	The dashed lines on the critical surface represent the intersection lines
	between these two kinds of the surface.
	The self-consistent surfaces under the critical surface are spurious.
	They have been shown due to a limit of the ability of our graphic software.
	}
\label{fig:self-consistent}
\end{figure}

\begin{figure}
	\includegraphics[clip]{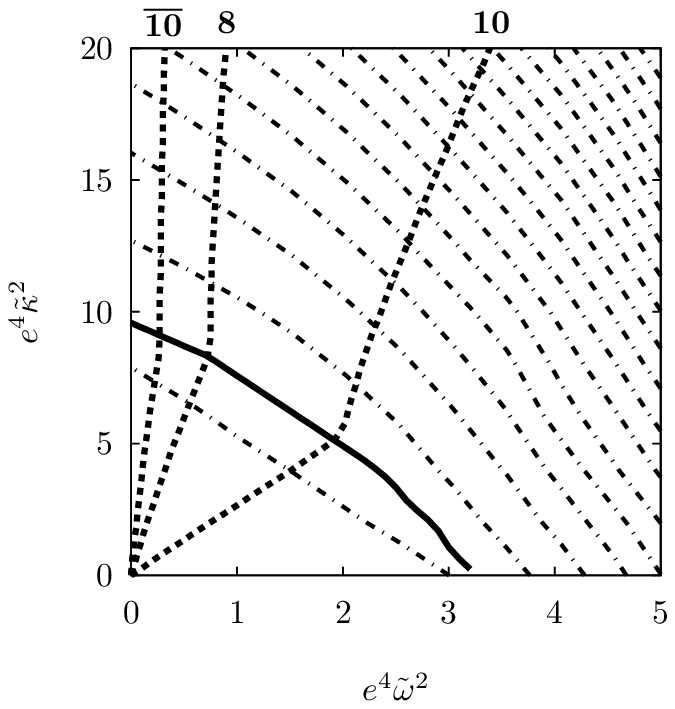}
	\caption{
	Intersection lines between the critical surface and the self-consistent surfaces
	for the multiplets ${\bf 8}$, ${\bf 10}$, and $\lbar{\bf 10}$.
	The dashed lines represent the intersection lines.
	The dash-dotted lines represent the contour lines of the critical surface
	with step $\Delta \tilde{m}_{\rm eff}^{2} = 2$.
	The solid line indicates the boundary curve in Fig.~\ref{fig:critical_surface}.
	}
\label{fig:critical_surface_contour}
\end{figure}

Since the number of the independent parameters is two,
the self-consistent solutions form surfaces for each multiplet in parameter space ${\cal M}$.
The surfaces are limited by the critical surface in Fig.~\ref{fig:critical_surface}.
We call the surfaces ``self-consistent surfaces'' and show these in Fig.~\ref{fig:self-consistent}.
In addition, Fig.~\ref{fig:critical_surface_contour} shows the intersection lines
between the critical surface and the self-consistent surfaces
on the contour map of the critical surface.
The self-consistent surface for ${\bf 10}$ is away from the others
because of Eqs.~(\ref{eq:wk2_8_10}).
Further the surface for $\lbar{\bf 10}$ is closer to $e^{4} \tilde{\omega}^{2} = 0$ plane
than the surface for ${\bf 8}$ because of Eqs.~(\ref{eq:wk2_8_10bar}).

These figures are useful for relating the profile functions
with multiplets ${\bf 8}$, ${\bf 10}$, and $\lbar{\bf 10}$.
For example, Fig.~(\ref{fig:self-consistent}) shows that
the profile functions in case A in Fig.~\ref{fig:near_critical} are excluded,
because point A is on $e^{4} \tilde{\kappa}^{2} = 0$ plane in ${\cal M}$ and
any self-consistent surfaces do not pass through this plane
except for the $\tilde{m}_{\rm eff}^{2}$ axis.
However the profile functions in the cases of R, B, and C can be self-consistent solutions.

Figure~\ref{fig:critical_surface_contour} shows that
if the self-consistent solution is evaluated near ${\cal S}_{H}$,
always $e^{4} \tilde{\kappa}^{2} > 4$.
Then the profile functions of ${\bf 8}$ and $\lbar{\bf 10}$ are similar
to that of case B in Fig.~\ref{fig:near_critical}, and 
the profile functions for ${\bf 10}$ are similar to that of case C.

\begin{figure}
	\includegraphics[clip]{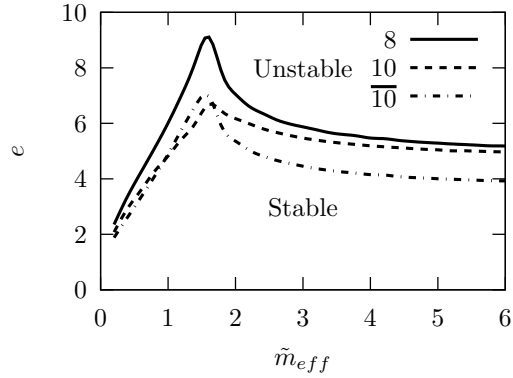}
	\caption{Critical curves for multiplets ${\bf 8}$, ${\bf 10}$, and $\lbar{\bf 10}$
		in parameter space $(e, \tilde{m}_{\rm eff})$.
        These curves separate the areas in which the soliton is stable or unstable.
        The horizontal axis $\tilde{m}_{\rm eff}$ is different from
		$m_{\rm eff}/F_{\pi} = m_{\rm eff}/(2 f_{\pi})$
		of Ref.~\cite{rf:Rajaraman86}.}
\label{fig:ccurve}
\end{figure}

Also in parameter space $(e, \tilde{m}_{\rm eff})$, there are curves
that separate the area of the stable soliton solutions and the area of the unstable ones in each multiplet.
We call the curves ``critical curves" and show these in Fig.~\ref{fig:ccurve}.
Every curve has an upward ledge at $\tilde{m}_{\rm eff} \sim 1.6$.
The left area of the ledge is restricted by Eq.~(\ref{eq:stcond1}), and
the right one by Eq.~(\ref{eq:stcond2}).
These curves correspond to the intersection lines
between the critical surface and the self-consistent surfaces in Fig.~\ref{fig:self-consistent}.
One can use this figure to decide whether the adjustable parameters $(e, f_{\pi}, m_{\pi}, m_{K})$
admit the stable soliton solution through Eq.~(\ref{eq:meff}).

The critical curve should reach $(e, \tilde{m}_{\rm eff}) = (0, 0)$ because of Eq.~(\ref{eq:stcond1}).
However, it is difficult to decide whether the solution of Eq.~(\ref{eq:eomF})
is stable for $\tilde{m}_{\rm eff} \sim 0$,
because the instability for $\tilde{m}_{\rm eff} \sim 0$ appears at the large radius $\rho \gg 1$.
Therefore, we show the curves only for $\tilde{m}_{\rm eff} \geq 0.2$.

For $\tilde{m}_{\rm eff} > 1.6$, the areas of the stable soliton become narrow,
because the moments of inertia $(\tilde{\alpha}^{2}, \tilde{\beta}^{2})$
are the decreasing functions of $\tilde{m}_{\rm eff}$ in our approach, and as a result
the stability condition~(\ref{eq:stcond2}) with Eqs.~(\ref{eq:omega}) and (\ref{eq:kappa})
becomes severe for parameter $e$.

From Eqs.~(\ref{eq:stcond1}), (\ref{eq:stcond2}), (\ref{eq:wk2_8_10}), and (\ref{eq:wk2_8_10bar}),
it is reasonable that the critical curve for ${\bf 8}$ is higher than those for ${\bf 10}$ and $\lbar{\bf 10}$
through all value of $\tilde{m}_{\rm eff}$.
On the other hand, from Eqs.~(\ref{eq:stcond1}), (\ref{eq:stcond2}), and (\ref{eq:wk2_10_10bar}),
it is not clear which critical curve for ${\bf 10}$ and $\lbar{\bf 10}$ is higher.
Indeed, the curves for ${\bf 10}$ and $\lbar{\bf 10}$ change their relative height
according to $\tilde{m}_{\rm eff}$.
In particular, the critical curve for ${\lbar{\bf 10}}$ is lower than that for ${\bf 10}$
in the area $\tilde{m}_{\rm eff} > 1.6$.
It is explained as follows.
The self-consistent solutions corresponding to the critical curves are obtained
near ${\cal S}_{H}$ in Fig.~\ref{fig:self-consistent}, and the solutions have $e^{4} \tilde{\kappa}^2 > 4$.
Therefore, condition~(\ref{eq:stcond2}) reaches $\rho = 0$.
For a large effective mass, condition~(\ref{eq:stcond2}) is effective at $\rho \sim 0$
because of Eq.~(\ref{eq:asymF}) and
it reduces to Eq.~(\ref{eq:stcond2a}) containing only $\tilde{\kappa}^{2}$.
Since $\tilde{\kappa}^{2}$ for $\lbar{\bf 10}$ is about three times larger
than that for the other multiplets in Eqs.~(\ref{eq:k2_8_10bar}) and (\ref{eq:k2_10_10bar}),
the area of the stable solitons for ${\lbar{\bf 10}}$ is narrower than those for ${\bf 8}$ and ${\bf 10}$.

The terms proportional to $\tilde{\kappa}^{2}$ in Eqs.~(\ref{eq:stcond2}) and (\ref{eq:stcond2a}) originate
from the term $\frac{e^{4}}{2} \tilde{\beta}^{2} \tilde{\kappa}^{2}$ in Eq.~(\ref{eq:hb}).
This term $\frac{e^{4}}{2} \tilde{\beta}^{2} \tilde{\kappa}^{2}$ 
is regarded as the coupling between the rotation into the strangeness direction and
the amplitude $\sin(F/2)$ in $\tilde{\beta}^{2}$.
The amplitude is the zero mode fluctuation around the hedgehog profile
in the flavor symmetry limit~\cite{rf:Callan88},
and it represents an intrinsic motion on the soliton.
If this term becomes large compared with $\tilde{M}_{0}$ in the mean field Hamiltonian~(\ref{eq:hb}),
the collective rotation and the intrinsic motion cannot dynamically separate and RRA fails.
That is a limit of the applicability of RRA pointed out from a general argument in Ref.~\cite{rf:Cohen04}.
In our approach, the influence of this coupling is dynamically included in the calculation of
the profile function through the mean field Hamiltonian~(\ref{eq:hb}). 
Therefore, condition~(\ref{eq:stcond2}) represents a more realistic limit so that
the coupling does not destroy the soliton itself.

\section{Baryon mass\label{sec:mass}}
With the profile function of the stable soliton,
the classical soliton mass~(\ref{eq:mcl}), the moments of inertia~(\ref{eq:alpha2}), (\ref{eq:beta2}),
and the symmetry breaking~(\ref{eq:gamma}) are evaluated.
Then the eigenvalue equation~(\ref{eq:eigeneq}) is numerically solved by the Yabu and Ando method,
and the baryon masses are obtained.
We introduce
\begin{equation}
	\Delta E^{\rm err} = \sqrt{
		\sum_{B \in {\bf 8},{\bf 10}} \left(\Delta E_{BN}-\Delta E^{exp}_{BN}\right)^{2}
	},
\label{eq:massdiff}
\end{equation}
where $\Delta E_{BN}$ is a difference in the predictive value of the mass
between the baryon $B$ and the nucleon $N$,
and $\Delta E^{exp}_{BN}$ is its experimental value.
An isospin multiplet is represented as a baryon $B$ in this formula,
because baryons in an isospin multiplet are described by the same soliton solution.
The quantity $\Delta E^{\rm err}$ measures an error of the predicted baryon mass splitting
for the multiplets ${\bf 8}$ and ${\bf 10}$.

The parameters $(e, \tilde{m}_{\rm eff})$ of the stable solitons are placed
below the critical curves in Fig.~\ref{fig:ccurve}.
A parameter set $(e, \tilde{m}_{\pi}, \tilde{m}_{K})$ corresponds
to the 12 points in the parameter space $(e, \tilde{m}_{\rm eff})$
according to Eq.~(\ref{eq:meff}) and Table~\ref{tb:qparam}.
While the Skyrme parameter $e$ determines the vertical positions of the points,
the masses $\tilde{m}_{\pi,K}$ give the horizontal positions and the spreads of the points.
We discuss only parameters $(e, \tilde{m}_{\pi}, \tilde{m}_{K})$
which admit the existence of the ${\bf 8}$, ${\bf 10}$, and $\lbar{\bf 10}$ baryons.
Since $(e, \tilde{m}_{\pi}, \tilde{m}_{K})$ are dimensionless,
the energy scale $(f_{\pi}/e)$ cannot be specified, and
the stability of the solitons is not sufficient
to determine the values of all parameters $(e, f_{\pi}, m_{\pi}, m_{K})$.
Using the degrees of freedom, we fit the $N-\Delta$ mass difference
or the absolute value of the $N$ mass to its experimental value.

Equation~(\ref{eq:EB}) gives the experimental value of the $N-\Delta$ mass difference and
the accurate baryon mass splitting (small $\Delta E^{\rm err}$).
The parameter set and the predicted baryon masses are shown as (1)
in Tables~\ref{tb:param} and \ref{tb:masslevel}, respectively.
Here, the value of $f_{\pi}$, $93$~(MeV), is given by hand.
A larger value of $f_{\pi}$ gives a slightly smaller value of $\Delta E^{\rm err}$,
but it leads to extremely larger baryon masses in proportion to $f_{\pi}$;
for example, $\Delta E^{\rm err} = 74$~(MeV) and $E_{N} = 5992$~(MeV)
for $e = 2.8$, $f_{\pi} = 186$~(MeV), $m_{\pi} = 106$~(MeV), and $m_{K} = 747$~(MeV).
Table~\ref{tb:set1} shows the dimensionless effective mass~($\tilde{m}_{\rm eff}$),
the classical mass~($M_{0}$), the moments of inertia~($\alpha^{2}$, $\beta^{2}$),
and the symmetry breaking~($\gamma$) for parameter set~(1).
In our approach, these quantities vary according to the baryon states:
$\tilde{m}_{\rm eff}$ characterizes each baryon state,
$M_{0}$ is the increasing function of $\tilde{m}_{\rm eff}$, and
($\alpha^{2}$, $\beta^{2}$, $\gamma$) are the decreasing functions.

However, Eq.~(\ref{eq:EB}) cannot give the experimental value of the $N$ mass and
the accurate mass splitting simultaneously in our approach.
In particular, the values of the baryon masses are large compared with the observed ones. 
That is a common phenomena in the Skyrme model~\cite{rf:Yabu88},
but the tendency is more severe in our approach.
The experimental values of the baryon masses lead to the smaller value of $f_{\pi}$
which is the energy scale in this model.
Therefore the value of the Skyrme parameter should be large
for the rotational energy to generate the mass splitting.
However, since the Skyrme parameter is restricted by the critical curves
for ${\bf 10}$ and $\lbar{\bf 10}$ in Fig.~\ref{fig:ccurve},
the magnitude of the mass splitting is not sufficiently large.

There is a more fundamental method~\cite{rf:Moussallam91,rf:Holzwarth93,rf:Holzwarth94,rf:Kim98} 
for the Skyrme model to reproduce the observed baryon masses.
In this method, the Casimir energy ($< 0$) due to the existence of the soliton is added to the baryon masses.
In RRA, the Casimir energy is the quantity of order $N_{c}^{0}$, and
the value is common to all baryon states.
Therefore the addition of this energy does not change the mass splitting, and
one can discuss the mass splitting and the values of the masses separately.
In our approach, the Casimir energy changes its value
according to the baryon states as well as the shape of the soliton
and contributes to the mass splitting too. 
Thus, our self-consistent procedure should include the effect of the Casimir energy
to treat the mass splitting and the masses themselves simultaneously.
However that is a complicated task to be examined in detail here,
because the simple analytic form of the Casimir energy is not known. 

Instead, we adopt a subtraction method~\cite{rf:Yabu88} to estimate the effect.
In this method, the unsubtracted mass formula~(\ref{eq:EB}) is replaced
by the subtracted one:
\begin{eqnarray}
	\tilde{E}_{B} &=& \tilde{M}_{0}
	+ \frac{e^4}{2} \left(\frac{1}{\tilde{\alpha}^2}-\frac{1}{\tilde{\beta}^2}\right) J (J+1)
	\nonumber \\ &&
	-\frac{3e^4}{8\tilde{\beta}^{2}} + \frac{e^4}{2\tilde{\beta}^{2}} ({\cal E}_{SB}-{\cal E}_{0}),
	\label{eq:EB_E0}
\end{eqnarray}
where the quantity ${\cal E}_{0}$ is the lowest eigenvalue of Eq.~(\ref{eq:YA})
corresponding to the vacuum-like state with $(I,J,Y,Y_{R}) = (0,0,0,0)$.
Equation~(\ref{eq:EB_E0}) improves the behavior of $\tilde{E}_{B}$
by removing the vacuum fluctuation energy according to the increase of the symmetry breaking and
reproduce the mass splitting accurately~\cite{rf:Yabu88}.

Table~\ref{tb:masslevel} shows the baryon masses calculated
by Eq.~(\ref{eq:EB_E0}) with parameter sets (2) and (3) given at Table~\ref{tb:param}.
Set (2) fits the $N-\Delta$ mass difference, and set (3) fits the $N$ mass.
Both the parameter sets give the accurate mass splitting.
In addition, Table~\ref{tb:set2} gives the values
of $\tilde{m}_{\rm eff}$, $M_{0}$, $\alpha^{2}$, $\beta^{2}$,
and $\gamma$ for set~(2), and Table~\ref{tb:set3} gives those for set~(3).

The deformation of the soliton reproduces the mass splitting accurately
for any parameter set given at Table~\ref{tb:param}, 
and it has the sizable effects on the ${\bf 8}$, ${\bf 10}$, and $\lbar{\bf 10}$ baryons masses
as seen from Tables~\ref{tb:set1}, \ref{tb:set2}, and \ref{tb:set3}.
However the mass splitting is caused by the different terms of the Hamiltonian
according to the parameters.
For example, the contributions of these terms to the $N-\Xi$ mass difference
are estimated at the difference in 
$\left(
M_{0}, \frac{1}{2} \alpha^{2} \omega^{2},
\frac{1}{2} \beta^{2} \kappa^{2},
\frac{3}{4} \gamma q_{B}
\right)$ calculated to $\Xi$ and $N$.
In RRA the $N-\Xi$ mass difference is dominated by the symmetry breaking term.
However, in our approach, it is distributed as follows:
$(147,  7, 60, 126)$~(MeV) for set~(1),
$( 51, 11, 72, 254)$~(MeV) for set~(2), and
$( 17, 14, 98, 201)$~(MeV) for set~(3).
Therefore the effects of the rotation and the symmetry breaking mix
through the deformation of the soliton each other,
and the sizes of the effects are large.

If the baryon masses are given by parameter set~(3),
the ${\bf 10}$ and $\lbar{\bf 10}$ baryons are affected obviously by the critical curve,
because the set corresponds to the points ($e = 6.17$, $\tilde{m}_{\rm eff} = 1.36 - 1.81$)
just below the critical curves for ${\bf 10}$ and $\lbar{\bf 10}$ in Fig.~\ref{fig:ccurve}.
Then, as mentioned in Sec.~\ref{subsec:multiplet_dep},
the profile functions of ${\bf 8}$ and $\lbar{\bf 10}$ are similar
to that of case B in Fig.~\ref{fig:near_critical}, and
the profile functions of ${\bf 10}$ are similar to that of case C.

\begin{table}
\caption{Parameter sets. ``Exp.'' denotes the experimental values.}
\begin{ruledtabular}
\begin{tabular}{lcccc}
Set & (1) & (2) & (3) & Exp.\\
\hline
$e$ & 3.00 & 3.40 & 6.05 & $-$ \\
$f_{\pi}$~(MeV) & 93 & 147 & 46 & 93 \\
$m_{\pi}$~(MeV) & 196 & 0 & 56 & 140 \\
$m_{K}$~(MeV) & 1042 & 616 & 551 & 496
\end{tabular}
\end{ruledtabular}
\label{tb:param}
\end{table}   

\begin{table}
\caption{Baryon mass differences from the nucleon mass
for parameter sets~(1), (2), and (3).
``Exp.'' denotes the experimental values.
$\Delta E^{\rm err}$ is defined by Eq.~(\ref{eq:massdiff}).
Only row of $N$ gives the absolute values of the nucleon mass.
Marks ``*'' denote the input values for the energy scales.
All units are (MeV).}
\begin{ruledtabular}
\begin{tabular}{lcccc}
Set & (1) & (2) & (3) & Exp.\\
\hline
$\Delta E^{\rm err}$ & 90 & 104  & 119 & $-$\\
${\bf 8}$\\
$N$ (abs.)		&3618 & 3483 & 939* & 939\\
$\Lambda$		& 187 &  177 & 154 & 183\\
$\Sigma$ 		& 314 & 310 & 270 & 256\\
$\Xi$ 			& 396 & 382 & 326 & 379\\
${\bf 10}$\\
$\Delta$		& 293* & 293* & 307 & 293\\
$\Sigma^{*}$	& 445 & 435 & 442 & 445\\
$\Xi^{*}$		& 570 & 554 & 551 & 595\\
$\Omega$		& 672 & 654 & 641 & 733\\
$\lbar{\bf 10}$\\
$\Theta^{+}$ 				& 708 & 483 & 441 & 601?\\
$N^{*}_{\lbar{10}}$ 		& 888 & 661 & 613 & ?\\
$\Sigma^{*}_{\lbar{10}}$ 	& 1022 & 800 & 746 & ?\\
$\Xi^{*}_{\lbar{10}}$ 		& 1059 & 865 & 819 & ?
\end{tabular}
\end{ruledtabular}
\label{tb:masslevel}
\end{table}

\begin{table}
\caption{Dimensionless effective mass $\tilde{m}_{\rm eff}$ and
parts of the Hamiltonian for the parameter set~(1):
classical mass $M_{0}$, moments of inertia ($\alpha^{2}$, $\beta^{2}$),
and symmetry breaking $\gamma$.}
\begin{ruledtabular}
\begin{tabular}{cccccc}
& & (MeV) & \multicolumn{2}{c}{($10^{-3}$/MeV)} & (MeV) \\
& $\tilde{m}_{\rm eff}$ & $M_{0}$ & $\alpha^{2}$ & $\beta^{2}$ & $\gamma$ \\
\hline
${\bf 8}$\\
$N$				& 2.60 & 2707 & 6.85 & 2.10 & 1628\\
$\Lambda$		& 2.92 & 2770 & 6.48 & 1.95 & 1393 \\
$\Sigma$		& 3.22 & 2828 & 6.19 & 1.84 & 1226\\
$\Xi$			& 3.35 & 2855 & 6.07 & 1.79 & 1159\\
${\bf 10}$\\
$\Delta$ 		& 2.89 & 2742 & 6.89 & 2.08 & 1531\\
$\Sigma^{*}$ 	& 3.07 & 2779 & 6.69 & 2.00 & 1406\\
$\Xi^{*}$ 		& 3.25 & 2813 & 6.51 & 1.93 & 1302\\
$\Omega$ 		& 3.42 & 2845 & 6.36 & 1.87 & 1216\\
$\lbar{\bf 10}$\\
$\Theta^{+}$ 				& 2.69 & 2647 & 7.90 & 2.45 & 2087\\
$N^{*}_{\lbar{10}}$ 		& 2.89 & 2681 & 7.63 & 2.34 & 1890\\
$\Sigma^{*}_{\lbar{10}}$ 	& 3.07 & 2714 & 7.42 & 2.25 & 1732\\
$\Xi^{*}_{\lbar{10}}$ 		& 3.25 & 2746 & 7.23 & 2.17 & 1602\\
\end{tabular}
\end{ruledtabular}
\label{tb:set1}
\end{table}

\begin{table}
\caption{Dimensionless effective mass and
parts of the Hamiltonian for the parameter set~(2).}
\begin{ruledtabular}
\begin{tabular}{cccccc}
& & (MeV) & \multicolumn{2}{c}{($10^{-3}$/MeV)} & (MeV) \\
& $\tilde{m}_{\rm eff}$ & $M_{0}$ & $\alpha^{2}$ & $\beta^{2}$ & $\gamma$ \\
\hline
${\bf 8}$\\
$N$				& 0.84 & 3200 & 5.71 & 2.08 & 1134\\
$\Lambda$		& 0.95 & 3221 & 5.33 & 1.91 & 990\\
$\Sigma$		& 1.05 & 3241 & 5.04 & 1.78 & 886\\
$\Xi$			& 1.10 & 3251 & 4.92 & 1.73 & 843\\
${\bf 10}$\\
$\Delta$ 		& 0.94 & 3199 & 5.88 & 2.15 & 1181\\
$\Sigma^{*}$ 	& 1.00 & 3211 & 5.65 & 2.04 & 1092\\
$\Xi^{*}$ 		& 1.07 & 3222 & 5.46 & 1.96 & 1018\\
$\Omega$ 		& 1.12 & 3234 & 5.30 & 1.89 & 955\\
$\lbar{\bf 10}$\\
$\Theta^{+}$ 				& 0.87 & 3174 & 7.23 & 2.74 & 1693\\
$N^{*}_{\lbar{10}}$ 		& 0.94 & 3180 & 6.88 & 2.58 & 1550\\
$\Sigma^{*}_{\lbar{10}}$ 	& 1.00 & 3187 & 6.59 & 2.46 & 1434\\
$\Xi^{*}_{\lbar{10}}$ 		& 1.07 & 3194 & 6.35 & 2.35 & 1338
\end{tabular}
\end{ruledtabular}
\label{tb:set2}
\end{table}

\begin{table}
\caption{Dimensionless effective mass and
parts of the Hamiltonian for the parameter set~(3).}
\begin{ruledtabular}
\begin{tabular}{cccccc}
& & (MeV) & \multicolumn{2}{c}{($10^{-3}$/MeV)} & (MeV) \\
& $\tilde{m}_{\rm eff}$ & $M_{0}$ & $\alpha^{2}$ & $\beta^{2}$ & $\gamma$ \\
\hline
${\bf 8}$\\
$N$			& 1.36 & 578 & 4.77 & 1.85 & 917 \\
$\Lambda$	& 1.54 & 584 & 4.4 & 1.67 & 776 \\
$\Sigma$	& 1.70 & 591 & 4.14 & 1.54 & 675 \\
$\Xi$		& 1.77 & 596 & 4.05 & 1.49 & 635 \\
${\bf 10}$\\
$\Delta$ 		& 1.52 & 606 & 5.31 & 2.05 & 1013 \\
$\Sigma^{*}$ 	& 1.63 & 621 & 5.1 & 1.94 & 902 \\
$\Xi^{*}$ 		& 1.72 & 640 & 4.97 & 1.84 & 808 \\
$\Omega$ 		& 1.81 & 660 & 4.88 & 1.76 & 731 \\
$\lbar{\bf 10}$\\
$\Theta^{+}$ 				& 1.41 & 613 & 6.87 & 2.84 & 1586 \\
$N^{*}_{\lbar{10}}$ 		& 1.52 & 619 & 6.49 & 2.66 & 1433 \\
$\Sigma^{*}_{\lbar{10}}$ 	& 1.62 & 627 & 6.20 & 2.52 & 1307 \\
$\Xi^{*}_{\lbar{10}}$ 		& 1.72 & 640 & 5.98 & 2.41 & 1198
\end{tabular}
\end{ruledtabular}
\label{tb:set3}
\end{table}

\section{Summary\label{sec:summary}}
In this paper, we have investigated the profile function of the SU(3) Skyrmion depending
on the octet, decuplet and antidecuplet baryon states.
The equations of motion for the profile function are given by the variation of the mean field Hamiltonian.
The Hamiltonian is the expectation value of the collective Hamiltonian operator for the baryon state
and depends on the profile function itself through the moments of inertia.
Thus, we should solve the equations of motion self-consistently.
As the result, the profile function is affected by not only the rotation of the Skyrmion
but also the flavor symmetry breaking.

The influence of the symmetry breaking on the profile function is represented
by an effective meson mass which varies according to the baryon states.
The effective mass in the rigid rotator approach is the pion mass and usually small.
In our approach, the effective mass can take a value about the kaon mass and,
the qualitatively different behavior of the soliton solution emerges.

In general, the rotation pushes the profile function out of the center of the soliton,
and the symmetry breaking (the effective meson mass) attracts the profile function.
For small effective mass, these effects are represented by the single mass scale
in the asymptotic form of the profile function.
Then the instability of the soliton appears as the spontaneous emission of the real meson
and restricts the parameter space of the self-consistent solutions.
That is similar to the result of the SU(2) Skyrmion.

For large effective mass, the influence of the rotation appears in the three cases.
First, the rapid spatial rotation leads to the large variation of the curvature ($F''$)
of the profile function at the intermediate radius.
Secondly, the rapid flavor rotation leads to the negative value of the curvature at the small radius.
Thirdly, the flavor rotation enhances the effect of the spatial rotation and
leads to the large variation of the curvature at the intermediate radius.
Although the first case is excluded by the self-consistent calculation,
the last two cases can be the self-consistent solutions and
restrict the parameter space.

The independent parameters of the self-consistent solution are
the Skyrme parameter and the effective meson mass.
There are areas of the independent parameters allowed for each multiplet.
The allowed value of the Skyrme parameter for the octet baryons is the largest 
for all value of the effective mass, and
those for the decuplet and antidecuplet baryons
change the relative size according to the effective mass. 
At the large effective mass, the allowed value of the Skyrme parameter for the antidecuplet baryons
is smaller than those for the octet and decuplet baryons.

The baryon masses are evaluated by the unsubtracted mass formula and the subtracted one, respectively.
Then, the deformation of the soliton reproduces the baryon mass splitting accurately
with both the mass formulas and has the sizable effects on the baryon masses.
Therefore the effects of the rotation and the symmetry breaking cannot separate clearly.

The subtracted mass formula can reproduce not only the mass splitting
but also the observed masses,
though the pion decay constant is too small.
Since the formula is inspired by the Casimir effect,
the Casimir energy should be investigated for our self-consistent procedure
to solve the problem of the small pion decay constant.
It remains as a matter to be researched further.

Other physical properties (e.g. magnetic moment, charge radius, etc\dots)
are affected by the deformation.
The study in this direction is in progress.


\end{document}